\documentclass[sigconf]{acmart}

\usepackage{booktabs} 
\renewcommand\footnotetextcopyrightpermission[1]{}
\usepackage{multirow}
\usepackage{subcaption}
\usepackage{caption}
\usepackage{setspace}
\usepackage{amsmath}
\usepackage[english]{babel}
\definecolor{mygray}{gray}{0.5}
\usepackage{float}
\usepackage{flushend}
\usepackage{mathtools}
\usepackage{soul}
\usepackage{arydshln}

\usepackage{enumitem}

\long\def\comment#1{}
\long\def\comments#1{}

\author{Yifan Qiao, Yingrui Yang, Haixin Lin, Tianbo Xiong, Xiyue Wang, Tao Yang}
\affiliation{%
  \institution{Department of Computer Science, University of California}
  \city{Santa Barbara}
  \state{California}
  \postcode{93106}
  \country{USA}
}

\setcopyright{none} 

\begin{document}

\title{ Dual Skipping Guidance for Document Retrieval  with Learned Sparse Representations}

\begin{abstract}

This paper proposes  a dual skipping guidance scheme with hybrid scoring to accelerate document  
retrieval that uses learned sparse representations while still delivering a good relevance.
This scheme  uses both lexical BM25 and learned neural term weights to bound and compose the rank score of
a candidate document separately for skipping and final ranking, and maintains two top-$k$ thresholds during inverted index
traversal. 
This paper evaluates time efficiency and ranking relevance of the proposed scheme  in searching MS MARCO TREC datasets.
 

\end{abstract}

\maketitle
\pagestyle{plain}

\section{Introduction}

\noindent


Document retrieval for searching a large dataset
often uses a simple additive linear ranking to select top-$k$ matched results before later-stage re-ranking.
Dynamic pruning techniques with document skipping
such as MaxScore~\cite{Turtle1995}, WAND~\cite{WAND},
and BlockMax WAND (BMW)~\cite{BMW}
can greatly speed up document retrieval 
by skipping low-scoring documents
that are unable to appear in the final top-$k$ list.
\comments{
As posting lists are often compressed in blocks, BMW~\cite{BMW}
divides each posting list into blocks,
stores the maximum score per block
and leverages such scores to skip unnecessary
decompression and inspections of posting blocks.
}
There are various further improvements to skip more documents effectively 
(e.g. ~\cite{
2003CIKM-Broder-SafeThreshold, 2013CandidateFiltering, 
2017WSDM-DAAT-SAAT,
Mallia2017VBMW, 
Tonellotto2018, Mackenzie2018, CIKM2020Additivity,
2018SIGIRKane, 2019ECIRMallia, 2019SIGIR-Petri,
2020CIKMComparison, 2012WSDMShan, khattab2020finding}), and
these studies  typically evaluate with BM25-based  term weights~\cite{Jones2000}.
Recently learned sparse representations have been developed to
compute term weights using  a neural model such as a transformer~\cite{Dai2020deepct, Bai2020SparTerm,
Formal2021SPLADE, Formal2021SPLADEV2,
mallia2021learning, Lin2021unicoil}. Together with document expansion (e.g.  ~\cite{Cheriton2019doct5query}),
document retrieval using learned sparse  representations can  deliver strong relevance results.
A downside is that
the retrieval time using learned sparse term weights is much slower than using BM25 weights  as discussed in  
~\cite{mallia2021learning, mackenzie2021wacky}. 

The main contribution of this paper is an add-on  control scheme
to provide dual-threshold skipping guidance to a retrieval algorithm,
and  to employ  a hybrid scoring with a linear combination of
BM25 and learned term weights for both skipping judgment and final ranking. 
The evaluation in  this paper with MS MARCO datasets  shows  that 
when applied  to variable-sized BMW~\cite{Mallia2017VBMW}, the proposed scheme can deliver a very competitive relevance while  
its retrieval speed is close to  BM25 retrieval with document expansion.
Both mean response time and 95th percentile time drop  significantly (e.g. varying from 1.5x to 4.3x) compared to the original baselines. 
Our scheme is significantly faster than a simple  threshold enlarging strategy~\cite{WAND, 2017WSDM-DAAT-SAAT}
in reaching a similar relevance level and can leverage such a strategy
for further  time  reduction when   $k$ is relatively large. 


\comments{
Such a technique was started  from DeepCT (Dai and Callan, 2019) and
recently, the DeepImpact work \cite{mallia2021learning} have  introduced a transformer model to produce sparse index scores
after using  document term expansion techniques based on Doc2query~\cite{X}.

Due to term expansion and distribution, pruning algorithms become less effective.~\cite{X}
}

\section{Background and Related Work}
\label{sect:background}

\noindent

The top-$k$ document retrieval problem identifies top ranked results in matching a query.
A document representation uses a feature vector to capture  the semantic of a document. 
If these vectors contain many  zeros, then such a representation is considered sparse.
For a large dataset, document retrieval often uses a
simple additive formula as the first stage of search and it computes the rank score of each document $d$ as: 
\[
 RankScore(d)= 
\sum_{t \in Q} w_t \cdot  w_{t,d},
\]
where $Q$ is the set  of all search terms,
$w_{t,d}$ is a weight contribution of  term $t$  in document $d$, 
and  $w_t$ is a document-independent or query-specific term weight. 
For the simplicity of presentation, assume that $w_{t,d}$ can be statically or dynamically scaled, and then we view  $w_t=1$.
An example of such formula is BM25~\cite{Jones2000} which is  widely used.

For a sparse representation, a retrieval algorithm  often  uses 
an \textit{inverted index} with a set of terms, and a \textit{document posting list} of each term
enumerating documents that contain such a term.
A posting record in a posting list  contains document ID and  its term weight.

{\bf Threshold-based skipping.} During the traversal of posting lists in document retrieval,
the  previous studies have advocated dynamic pruning strategies 
to skip low-scoring documents, which cannot appear on the final top-$k$ list
~\cite{WAND,strohman2007efficient}.
\comments{
Some information of a posting block $p$
can be accessible without decompression, and such information contains
the maximum weighted term feature score among all documents
and the maximum document ID in this block, denoted  as $BlockMax(p)$ and  $MaxDocID(p)$.
}
To skip the scoring of any document, a pruning  strategy computes the upper bound rank score of a candidate document $d$,
called $Bound(d)$. Namely this bound value  satifies $RankScore(d) \le Bound(d)$. 

If $Bound(d) < \Theta$ where $\Theta$ is a minimum rank score of documents in the top final $k$ list, the document can be skipped. 
For example, WAND~\cite{WAND} uses the maximum term weights of documents  of each posting list to determine the rank score upper bound of
a pivot document while BMW~\cite{BMW} and its variants (e.g. ~\cite{Mallia2017VBMW}) optimize WAND 
using  block-based maximum weights to compute the score upper bounds. MaxScore~\cite{Turtle1995} compares the
weight contribution sum upperbound of non-essential terms with the top-$k$ threshold to guide the partitioning of query terms.
In general, dynamic pruning with document skipping is   often used 
together with the document-at-a-time  or  term-at-a-time  traversal strategy~\cite{WAND,strohman2007efficient,Turtle1995,BMW}.

The above skipping is considered to be rank-safe up to $k$  in the sense that  the top-$k$ documents produced are ranked correctly.
Previous work has also pursued  a ``rank-unsafe'' skipping strategy by  deliberately over-estimating the current top-$k$ threshold 
by a factor of $F$~\cite{WAND, 2012SIGIR-SafeThreshold-Macdonald, 2013WSDM-SafeThreshold-Macdonald, 2017WSDM-DAAT-SAAT}.
There are also related strategies to obtain an accurate  top-$k$ threshold earlier, 
e.g. ~\cite{TwoTierBMW, 2019SIGIR-Petri, 2020CIKMComparison, 2019CIKMYafay,Shao2021SIGIR}.
While we can benefit from these studies, this paper does not study them because  they represent orthogonal optimizations. 
\comments{
Each block stores the maximum feature score per block, and 
the upper rank score of a document is bounded by the sum of the maximum block score of the blocks in which this document
may reside, and if such an estimated rank score bound is smaller than the top-$k$ threshold, the corresponding document can be
skipped for evaluation.

}

{\bf Learned sparse representations.}
Earlier sparse representation studies are conducted  in \cite{Zamani2018SNRM},
DeepCT~\cite{Dai2020deepct}, and SparTerm~\cite{Bai2020SparTerm}. 
\comments{
\citet{Dai2020deepct} learns  contextualized term weights to replace TF-IDF weights.  
}
Recent work on this subject includes 
SPLADE~\cite{Formal2021SPLADE, Formal2021SPLADEV2} learning token importance for  document expansion with sparsity control. 
DeepImpact~\cite{mallia2021learning} learns neural term weights on documents expanded by DocT5Query~\cite{Cheriton2019doct5query}. 
Similarly, uniCOIL~\citep{Lin2021unicoil} extends the work of COIL~\citep{2021NAACL-Gao-COIL} for contextualized term weights. 
Document retrieval with term weights learned from a transformer has been found slow in ~\cite{mallia2021learning, mackenzie2021wacky}. 
Mallia  et al. ~\cite{mallia2021learning} states that the MaxScore retrieval algorithm does not efficiently explpoit the DeepImpact scores.
Given ``wacky weights'' generated by a transformer affecting opportunities of document skipping during retrieval,
Mackenzie et al.~\cite{mackenzie2021wacky} advocated ranking approximation  with score-at-a-time  traversal. 

In this paper, we still focus on document-at-a-time retrieval, and propose a complementary scheme to accelerate  retrieval with dual-threshold skipping when using  a learned sparse 
representation while our design 
intends to preserve or even enhance the relevance.
Our skipping and final ranking adopts a hybrid formula to bound and combine rank scores based on BM25 weights and  learned term weights.
That is motivated by the recent studies in composing lexical and neural models in re-ranking~\cite{Yang2021WSDM-BECR}
and in combining scores from sparse retrieval and  dense retrieval~\cite{Lin2021tctcolbert, gao2021complementing, ma2021replication}.
We choose VBMW~\cite{Mallia2017VBMW}  to demonstrate our scheme because
VBMW is generally acknowledged to represent the state of the art~\cite{mackenzie2021wacky} for many cases.  
MaxScore could be a better choice for larger values of $k$ and for long queries~\cite{2019ECIRMallia}
and our technique could be  applicable to   MaxScore which uses threshold-based skipping, which is in our future work.

\section{Retrieval with Dual Guidance}
\label{sect:method}

\noindent

\begin{figure}[h]
    \centering
    \includegraphics[width=1\linewidth]{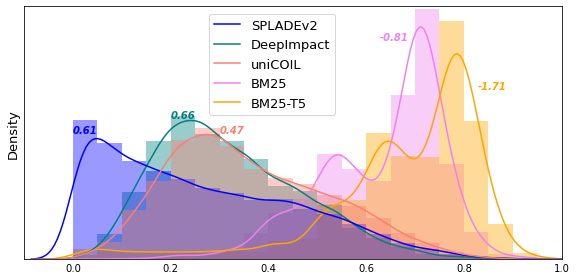}
    \caption{Distribution of BM25 and learned weights marked with distribution skewness.}
    \label{fig:distribution}
\end{figure}

Figure~\ref{fig:distribution} plots the 
min-max scaled distribution of term weights from BM25, BM25-T5, uniCOIL, DeepImpact, and SPLADEv2 of MS MARCO passages respectively.
We refer the BM25 scores calculated after DocT5Query expansion~\cite{Cheriton2019doct5query} as BM25-T5.
This figure shows that  BM25 weights and BM25-T5 weights  are 
left skewed while the weights from all learned sparse representations 
are skewed to the right. 
\comments{
A right skewed distribution indicates that there are more outliers with big scores,
and   for BMW/VBMW handling such a distribution, the maximum score  of a posting block  could be often less representative for this block.
Then this inaccuracy affects the chance  of skipping which depends on the accuracy of upper bound estimation.
}
An earlier study by Petri et al.  ~\cite{2013MoffatMagicWAND}
shows that the choice of the ranking score contribution fomula and their distribution have an impact 
on the effectiveness of index skipping during retrieval.
From that, one  can conjecture that the distribution right-skewness of learned weights in all three models
may be correlated to  their  slowness of  query processing compared to BM25 and BM25-T5 with a left-skewed distribution. 

As shown later in Section~\ref{sect:evaluation}, BM25-T5 does skip more documents during retrieval compared to uniCOIL, and
while a learned sparse representation performs well in terms of NDCG or MRR relevance numbers, BM25-T5 can still deliver a decent recall ratio
especially with large $k$ values. 
Our idea is that BM25-T5 weights 
can still be valuable  to augment a retriever with learned weights and 
guide skipping. With this in mind, our design considerations are listed below.

\begin{itemize} 
[leftmargin=*]

\item We treat  rank scoring for document skipping differently from the final result ranking.
Namely we keep two top-$k$ rank score thresholds during index traversal. One is based on BM25 weights, another is based on
the learned weights.
To accomplish that, 
each posting record in the inverted index contains two weights for each term in a document.
A retriever can maintain two queues for active top-$k$ results based on the above two weight types.
\item Bound estimation for skipping can be influenced by both BM25 and learned weights. 
Since we maintain two thresholds, we can use one threshold  to skip based on BM25 or BM25-influenced score as discussed below,
and another  to remove  documents  that will not appear in the final top-$k$  list. 

\item The above BM25-guided pruning is not rank-safe because it may skip some documents which are supposed to be included
in top-$k$ when strictly following the learned weight model.  
While rank-safeness is not a hard requirement, 
we plan to use two strategies to reduce the chance of   ``unsafe'' skipping.
One is to use a linear combination of both BM25 and learn weights in scoring  to compare  against a skipping threshold.
In this way, the impact of BM25 weights in skipping judgment is adjustable.
Since there are two top-$k$ queues maintained, 
our second strategy is to  study view consistency between two queues, which  can  improve skipping safeness.
We can also adopt the threshold over-estimation strategy~\cite{WAND, 2012SIGIR-SafeThreshold-Macdonald,2013WSDM-SafeThreshold-Macdonald,2017WSDM-DAAT-SAAT} 
that enlarges a skipping threshold by a factor to improve speed as long as it does not hurt relevance. 
\item Our second goal is to retain or improve relevance.
Since rank scores based on BM25 weights  and  learned weights are handily available,
the final rank score can be composed by a linear combination of these two scores, 
which provides  an opportunity of further relevance improvement with no extra cost.

\end{itemize}

\begin{figure}[htbp]
\begin{center}
  \includegraphics[width=\columnwidth]{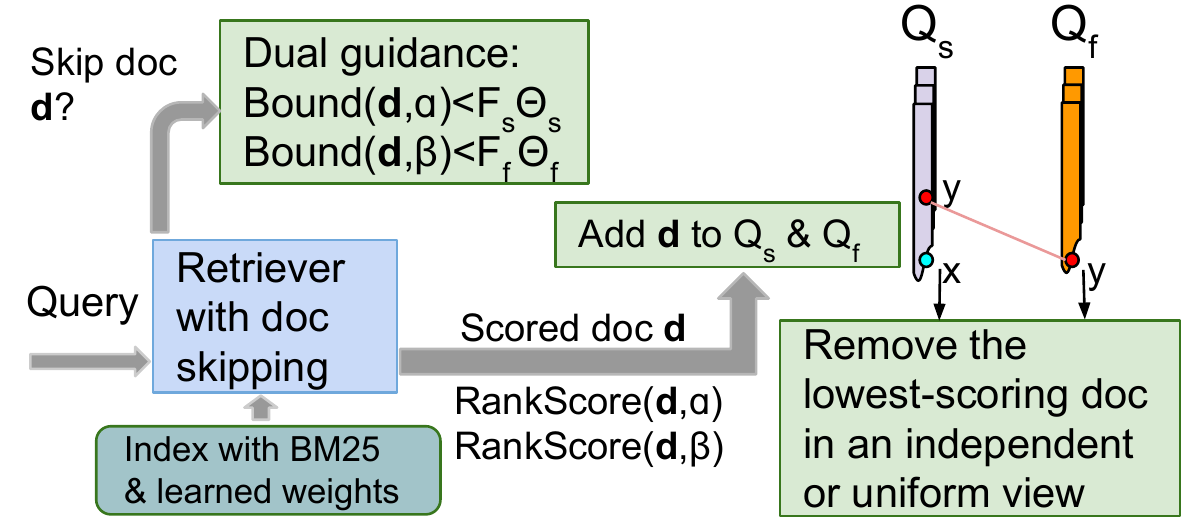}
\end{center}
  \caption{ Dual-threshold guidance  with hybrid scoring. ``s'' stands for skip-oriented scores, and ``f'' stands for final rank scores. $\Theta$ is the top-$k$ threshold, and $F$ is the over-estimating factor of $\Theta$.}
  \label{fig:guidedskip}
\end{figure}
Following the above discussion, we propose the 
following scheme with  dual-threshold skipping guidance and  hybrid scoring called DTHS.
Figure~\ref{fig:guidedskip} illustrates the control flow of DTHS and the rest of this section explains
each component of this figure in details. 
Given a retriever which uses threshold-based  skipping, we extend it  to 
use both BM25-based term weights and  learned weights, compute two score upper bounds and two rank scores of a candidate document,
and  consults two thresholds for document skipping. The details of control guidance imposed  to this retriever are described as follows.

\begin{itemize} 
[leftmargin=*]

\item  
When the underlying retriever computes the rank score upper bound of a  document for skipping judgment
based on the additive formula in Section~\ref{sect:background}, 
it needs to be extended to compute the following two bounds.
Let $Bound_B(d)$ be the estimated maximum rank score for document $d$ using BM25.
Let $Bound_L(d)$ be the estimated maximum rank score for $d$ using learned weights.
 A linear combination of these two estimated bounds will be used as the bound for skipping judgment: 
\[
Bound(d, \alpha) = \alpha Bound_B(d)  +(1-\alpha) Bound_L(d) 
\]
where $0\leq \alpha \leq  1$.
A large $\alpha$ value such as 1  means the skipping condition is mainly based on BM25 weights,
while  a smaller $\alpha$ value means  skipping is mainly based on learned weights.

\item Two rank scores are computed for each document $d$: $RankScore_B(d)$ and  $RankScore_L(d)$ based on 
 BM25 weights and learned weights, respectively.
We use the following linear combination as  the final score of document $d$ using parameter $\beta$.
\[
RankScore(d, \beta) =\beta RankScore_B(d)  +(1-\beta) RankScore_L(d)
\]
where $0\leq \beta \leq 1$.  If $\beta$ is 1, the final scoring purely follows learned weights. A linear combination may boost the relevance.
If $\alpha= \beta$, it means skipping uses the same scoring formula as the final ranking to guide pruning, and the top-$k$ retrieval algorithm is safe.

\item We maintain two separate queues: Queue $Q_s$ for the documents that have the $k$ largest  skip-oriented scores using  $RankScore(d, \alpha)$,
and Queue $Q_f$ for the documents with the $k$ largest final rank scores using $RankScore(d, \beta)$.
Top-$k$  threshold $\Theta_s$ is updated based on Queue $Q_s$ while
top-$k$ threshold  $\Theta_f$  is updated based on Queue $Q_f$. 




\item We have two options of making a skipping judgment where $F_s$ and $F_f$ are over-estimation factors:
\begin{itemize} 
[leftmargin=*]
\item {\bf Single-threshold  skipping (ST)}.
  If $Bound(d, \alpha) < F_s \Theta_s$, then scoring of document $d$ is skipped.
\item {\bf Dual-threshold  skipping (DT)}.
If $Bound(d, \alpha) < F_s \Theta_s$ or  $Bound(d, \beta) < F_f \Theta_f$, then scoring of document $d$ is skipped.
\end{itemize}

\item When the detailed scoring of document $d$ is not skipped,
this document  is  added to both queues. 
One document is removed from each queue to maintain its size as $k$.
As shown in Figure~\ref{fig:guidedskip},
let   document $x$ be the lowest scoring document in  $Q_s$ 
Let   document $y$ be the lowest scoring document in  $Q_f$.
If  $x = y$, we can just remove $x$ from both queues. 
When $x \neq y$, there are two options: 
\begin{itemize}
\item {\bf Independent view: }
The lowest-scoring document in each queue is removed separately without inter-queue coordination. 
This option allows different top-$k$ documents between $Q_s$ and $Q_f$ be maintained  
so  that $Q_s$ is more accurately matching the skipping condition regulated by $RankScore(x,\alpha)$ formula.
If removing document $x$ from $Q_s$ is a mistake because  its relevance is actually high based on the learned weights, 
since such a document is  still kept in $Q_f$, this document can still  appear in the final top-$k$ list.

\item {\bf Uniform view: } We remove $y$ from both queues, and in this way, two queues always contain the same  document sets.
This design option improves the pruning safeness.
Since  document $y$ will not appear in the final top-$k$ at the end,
keeping $y$ in $Q_s$ is unsafe. By removing $y$ from both queues makes two queues maintain a uniform  view of
what should be removed and kept.
\end{itemize}

\item At the end of retrieval, $Q_f$  outputs top-$k$ documents based on the combined  rank scores $RankScore(x,\beta)$. 

\end{itemize}

\section{Evaluation}
\label{sect:evaluation}




\noindent
\textbf{Setting and metrics.} 
Our evaluation uses
the MS MARCO document and passage collections for retrieval and ranking~\cite{Craswell2020OverviewOT, Campos2016MSMARCO}. The contents in the document collections are segmented during indexing and re-grouped after retrieval using ``max-passage'' strategy following~\cite{pyserini}.
There are 8.8M passages with an average length of 55 words, and 3.2M documents with an average length of 1131 words.
The Dev query set for passage and document ranking has 6980 and 5193  queries respectively  with about  one judgment label per query. 
Each of the passage/document  ranking task of TREC Deep Learning (DL) 2019 and 2020 tracks provides
a set of queries with many judgement labels per query.

In producing an inverted index, all words use lower case letters. 
The stand-alone BM25 and BM25-T5 indices reported in the following tables use  the BERT's Word Piece tokenizer.
For learned representations, the DeepImpact index  uses  a  tokenizer called nltk \cite{bird2009natural}
while uniCOIL and SPLADEv2 use the BERT's Word Piece tokenizer.
When BM25 is used with a learned representation for DTHS, their tokenization needs to be consistent.
For example, BM25 in uniCOIL/DTHS is computed for tokens based on the Word Piece tokenizer, 
while BM25 in DeepImpact/DTHS follows the nltk tokenizer. 
The index compression uses  SIMD-BP128~\cite{2015Lemire}, following~\cite{2019ECIRMallia}.
We apply VBMW~\cite{Mallia2017VBMW} with variable-sized blocks (the average block size is 1024 posting records). 

Our implementation uses C++, leveraging block partitioning code from ~\cite{Mallia2017VBMW}, and 
is compiled with GCC 10.2.0 and -Ofast optimization flag, 
running as a single thread on a Linux server with Intel i5-8259U 2.3GHz  and 32GB memory.

For MS MARCO Dev set, we report the relevance in terms of 
mean reciprocal rank (MRR@10 on passages and MRR@100 on documents), following the official leader-board standard.
One reason to choose MRR instead of
using normalized discounted cumulative gain (NDCG)~\cite{NDCG}  is
because  such a set has about one judgment label per query, 
which is too sparse to use NDCG.
For TREC DL test sets, we report  normalized discounted cumulative 
gain (NDCG@10)~\cite{NDCG}, following the common practice of the previous 
work~\cite{mallia2021learning,2021NAACL-Gao-COIL,gao2021complementing,Formal2021SPLADEV2}. NDCG is approrpiate for 
 DL test sets  because they have many judgement lables per query. 
We also report the 
recall ratio which is the percentage of relevant-labeled results appeared in the final top-$k$ results.

Before timing queries, all compressed posting lists and metadata for tested queries are pre-loaded into memory, 
following the same assumption in \cite{khattab2020finding, Mallia2017VBMW}.
Retrieval mean response times (MRT) are reported in milliseconds.
The 95th percentile time (95T)  is reported within parentheses in the tables below,
corresponding  to the time occurring in the 95th percentile and  called tail latency in ~\cite{Mackenzie2018}.
For all of our experiments, we perform pairwise t-test on the relevance between proposed method and corresponding baselines, no statistically significant degradation is observed at 95\% confidence level.


\begin{table}[htbp]
    \scriptsize

    \centering
        \caption{MRR(@10 for passages and @100 for documents), NDCG@10,  and mean/95th percentile time in milliseconds of different methods when $k$=1,000.}
\label{tab:overall}
        
        \resizebox{1.05\columnwidth}{!}{%
    \begin{tabular}{l|rr|rr|rr}
    \hline
     & \multicolumn{2}{c|}{\bf{Dev}} & \multicolumn{2}{c|}{\bf{TREC DL'19}} & \multicolumn{2}{c}{\bf{TREC DL'20}} \\
     \bf{Methods} & \bf{MRR} & \bf{Time} & \bf{NDCG} & \bf{Time} & \bf{NDCG} & \bf{Time}  \\ 
     
     \hline

     \multicolumn{7}{c}{\bf{MS MARCO Passages }} \\
     
     \hline
     BM25 & 0.172 & 17(42) & 0.425 & 52(140) & 0.453 & 54(157) \\
     BM25-T5 & 0.277 & 33(70) & 0.579 & 75(166) & 0.629 & 74(179) \\
     uniCOIL & 0.347 & 50(132) & 0.703 & 229(720) & 0.675 & 240(859) \\
     uniCOIL, $F$=1.7 & 0.346 & 30(74) & 0.703 & 73(183) & 0.675 &  75(234) \\
     uniCOIL, $F$=1.9 & 0.345 & \textbf{23(56)} & 0.695 & 55(124) & 0.672 & 60(172) \\
     DeepImpact & 0.328 & 71(166) & 0.695 & 235(751) & 0.628 & 243(855)  \\
     SPLADEv2 & 0.353 & 1373(2997) & 0.729 & 1513(3461) & 0.714 & 1400(3191) \\
     \hline
     uniCOIL/DTHS & 0.356 & 32(76) & 0.707 & 83(169) & 0.685 & 77(198)  \\
     + $F_s$=1.3,$F_f$=1 & 0.353 & 24(57) & 0.702 & \textbf{48(101)} & 0.675 & \textbf{46(110)} \\ 
     DeepImpact/DTHS & 0.344 & 45(96) & 0.710 & 108(302) & 0.675 & 111(261) \\
     SPLADEv2/DTHS & \textbf{0.362} & 896(1967) & \textbf{0.735} & 1059(2320) & \textbf{0.714} & 953(2067) \\

     \hline
     
     \multicolumn{7}{c}{\bf{MS MARCO Documents}} \\
     
     \hline
     
     BM25 & 0.203 & 165(351) & 0.446 & 171(373) & 0.451 & 169(422)  \\
     BM25-T5 & 0.303 & 205(435) & 0.559 & 179(374) & 0.561 & 174(408) \\
     uniCOIL & 0.353 & 494(1436) & 0.641 & 501(1461) & 0.601 & 512(2056) \\
    uniCOIL, $F$=1.5 & 0.352 & 207(525) & 0.641 & 203(531) & 0.601 & 203(609) \\
     uniCOIL, $F$=1.7 & 0.351 & 154(348) & 0.637 & 158(350) & 0.601 & 153(430) \\
     \hline
     uniCOIL/DTHS & \textbf{0.373} & 255(571) & \textbf{0.670} & 199(459) & \textbf{0.619} & 193(483) \\
     + $F_s$=1.3,$F_f$=1 & 0.371 & 173(385) & 0.670 & 127(282) & 0.619 & 125(304)  \\
     + $F_s$=1.5,$F_f$=1 & 0.367 & \textbf{130(284)} & 0.669 & \textbf{115(240)} & 0.619 & \textbf{113(269)}  \\ 
     \hline
    \end{tabular}
    }
\end{table}

\noindent
{\bf Relevance and time efficiency 
for retrieving top 1,000 results}. 
Table~\ref{tab:overall} reports performance on MS MARCO passage or document collections using VBMW.  
The DeepImpact, uniCOIL and SPLADEv2 weights  are based on data from Pyserini~\cite{pyserini},
after expansion with  DocT5Query. Default DTHS is configured with independent-view queue management, dual-thread (DT) skipping, $\alpha=0.9$, 
 $\beta=0.2$, $F_s=F_f=1$ without threshold over-estimation.

\comments{ 
 UniCOIL and SPLADEv2 require additional query encoding to calculate the weights of the query words during query processing. 
 This part of time is not included in Table~\ref{tab:overall}.
}
For MS MARCO passages, DTHS is compared with original ranking when using uniCOIL weights,
and  also with two  VBMW baselines using threshold over-estimation  by $F$ where $F$= 1.7  and 1.9.
The relevance of DTHS slightly outperforms original uniCOIL ranking on all test sets mainly indicating the hybrid scoring is effective. 
The over-estimation threshold is useful to reduce the latency while it does bring down the relevance number also.
The speedup in terms of MRT and 95T are 1.6x and 1.7x on Dev set. 
For DL'19 and DL'20 with a longer average query length, DTHS is 2.8x and 3.1x faster in terms of MRT and 4.3x and 4.3x faster in terms of 95T. 
Compared to simple  threshold enlarging (uniCOIL, $F=1.7$), DTHS has a visibly better relevance.
DTHS with $F_s=1.3$ is similar to uniCOIL with $F=1.7$ in  relevance  but it is 1.3x and 1.3x faster in MRT and 95T for Dev set,
and 1.5x and 1.8x faster for DL'19. 

DTHS also does well with DeepImpact and SPLADEv2 weights.
The relevance for DTHS remains comparable to the original DeepImpact and SPLADEv2 ranking
while MRT and 95T are over 1.5x faster on Dev set compared to original DeepImpact and SPLADEv2.  
 
The lower portion of Table~\ref{tab:overall} reports MRR@100  for the Dev set and
NDCG@10 for DL'19 and DL'20 in using BM25 and uniCOIL weights for MS MARCO document ranking when $k=1,000$.
The takeaways from these results are similar. 

\noindent
{\bf Index space cost.} The index size of BM25, BM25-T5, uniCOIL, DeepImpact, and SPLADv2 is   0.9, 1.4, 1.5, and 4.5, respectively  in GB for passages.
DTHS's size is  2, 2.2, and 6.7 in GB with these 3 learned models for passages.
The index size of BM25, BM25-T5, and uniCOIL for documents are 6.4, 7.1, 6.9 in GB while   
DTHS has  10.3GB with uniCOIL for documents. There is a modest increase of index space using DTHS because of hosting extra BM25 weights.

\comments{
\begin{table}[tpbh]
    \scriptsize
    \centering
        \caption{Recall.\label{tab:recall}}
\label{tab:smallk}
    \begin{tabular}{l|rrr|rr}
    \hline 
    
    \hline
     \bf{k} & \bf{BM25-T5} & \bf{UniCOIL} & \bf{w/ TH 1.7} & \bf{UniCOIL (DTHS)} & \bf{w/ TH 1.3}  \\ \hline
     10 & 0.5135(17) & 0.6101(22) &  & 0.6003(14) & \\
     20 & 0.6135(18) & 0.7067(23) &  & 0.6959(15) & \\
     200 & 0.8484(25) & 0.8992(30) & & 0.8971(24) & \\
     1000 & 0.9363(34) & 0.9527(50) & & 0.9530(32) & \\
     \hline
     
     \hline
    \end{tabular}
\end{table}
}

\begin{table}[tpbh]
    \centering

        \caption{MRR(Recall@k) and  MRT(95T) on the Dev query set.\label{tab:recall}}
    \setlength\tabcolsep{3pt}
\label{tab:smallk}

\resizebox{1.04\columnwidth}{!}{%
    \begin{tabular}{l|rr|rr|rr|rr}
    \hline 
     & \multicolumn{2}{c|}{\bf k=10} &\multicolumn{2}{c|}{\bf k=20}& \multicolumn{2}{c|}{\bf k=200} & \multicolumn{2}{c}{\bf k=1000} \\
      & MRR(R.) & Time & MRR(R.) & Time & MRR(R.) & Time  & MRR(R.) & Time \\ \hline
      \multicolumn{9}{c}{\textbf{MS MARCO Passages Dev}} \\ \hline
     BM25-T5 & .277(.513) & 17(42) & .277(.614) & 18(44) & .277(.848) & 25(54) & .277(.936) & 33(70)  \\
    \hline 
     uniCOIL & .347(.610) & 22(52) & .347(.707) & 23(55) & .347(.899) & 30(71) & .347(.953) & 50(132) \\
      + $F$=1.1 & .346(.606) & 21(54) & .346(.702) & 22(55) & .347(.899) & 29(70) & .347(.953) & 44(110) \\
      + $F$=1.3 & .322(.549) & 17(46) & .336(.655) & 19(50) & .346(.886) & 27(68) & .347(.953) & 36(91) \\
      + $F$=1.7 & .236(.350) & 10(34) & .276(.462) & 12(39) & .340(.796) & 19(49) & .346(.919) & 30(74) \\
    \hline 
     DTHS & .350(.600) & 14(38) &  .353(.696) & 15(42) & .355(.897) & 24(63) & .356(.957) & 32(76) \\
      + $F_s$=1.01 & .348(.596) & 13(38) & .352(.694) & 15(40) & .355(.894) & 24(62) & .356(.957) & 31(74) \\
      + $F_s$=1.1 & .287(.448) & 11(35) & .313(.559) & 12(36) & .352(.833) & 19(54) & .355(.933) & 28(67) \\
      + $F_s$=1.3 & .217(.315) & 7(30) & .260(.420) & 9(34) & .339(.751) & 17(49) & .353(.897) & 24(57) \\

     \hline
     
           \multicolumn{9}{c}{\textbf{MS MARCO Documents Dev}} \\ \hline
     BM25-T5 & .291(.571) & 79(196) & .298(.673) & 87(208) & .303(.883) & 128(295) & .303(.925) & 205(435)  \\
    \hline 
     uniCOIL & .342(.640) & 213(587) & .349(.745) & 230(651) & .353(.918) & 334(1000) & .353(.943) & 494(1436) \\
      + $F$=1.1 & .331(.562) & 142(337) & .344(.669) & 159(387) & .353(.889) & 244(598) & .353(.943) & 330(816) \\
      + $F$=1.3 & .310(.506) & 100(210) & .332(.618) & 150(322) & .352(.878) & 183(435) & .353(.940) & 255(604) \\
      + $F$=1.5 & .255(.390) & 75(158) & .291(.504) & 89(173) & .347(.828) & 135(280) & .352(.924) & 199(459) \\
    \hline 
     DTHS & .349(.573) & 102(227) & .363(.678) & 114(257) & .373(.887) & 181(418) & .373(.942) & 255(571) \\
      + $F_s$=1.01 & .348(.571) & 99(220) & .362(.676) & 110(254) & .373(.887) & 176(407) & .373(.942) & 252(564) \\
      + $F_s$=1.1 & .333(.533) & 82(173) & .351(.640) & 96(219) & .371(.872) & 150(340) & .372(.937) & 221(496) \\
      + $F_s$=1.3 & .267(.397) & 62(129) & .303(.509) & 69(142) & .362(.819) & 111(236) & .371(.916) & 173(385) \\

     \hline
    \end{tabular}
    }
\end{table}

    

     

\noindent
{\bf Performance when $k$ varies.} 
Table~\ref{tab:smallk}
compares relevance and query processing time of several methods when using uniCOIL weights and  $k=$  10, 20, 200, and 1,000 
for MS MARCO passages Dev set and MS MARCO documents Dev set.
The 95th percentile time  is reported in the parentheses next to the mean query latency number in the tables while
the recall ratio is reported in the parentheses next to the MRR number.
This recall ratio is the percentage of relevant-labeled results appeared in the final top-$k$ results.
For MS MARCO passages, MRR@10 is reported for different $k$  values.
For MS MARCO documents, MRR@100 number is reported when $k=$ 200 and 1000,
and MRR@k number is reported when $k=10$ and 20.

In general, DTHS performs well in relevance, MRT, and  95T for smaller $k$ values with a takeway similar as the case 
of $k=1,000$ studied above. 
For example, with $k=10$, DTHS performs similarly as original uniCOIL in relevance while it is 1.6x and 1.4x faster in MRT and 95T on passages Dev set. 
For small $k$ values such as $k=10$, simple threshold over-estimation becomes ineffective for both uniCOIL and DTHS because 
relevance drops significantly and  a small enlarging factor yields limited or no time reduction. 
For documents Dev set, the results and observations are similar.


\begin{table}[tpbh]
\scriptsize
    \centering
        \caption{ Characteristics of  design options on DL'19 passages}
        
        \setlength\tabcolsep{3pt}
        
\label{tab:options}
\resizebox{1.05\columnwidth}{!}{%
    \begin{tabular}{lrrrrr}
    \hline
     \bf{Methods} & \bf{Time} & \bf{NDCG} & \bf{Overlap} & \bf{BLoad} & \bf{\# Eval}  \\ 
     \hline
     BM25-T5 & 75(166) & 0.579 & 48.36\% & 69.33\% & 157072 \\
     uniCOIL & 229(720) & 0.703 & 100\% & 95.21\% & 643842 \\
     uniCOIL, $F$=1.7 & 73(183) & 0.703 & 50.51\% & 74.22\% & 110221 \\
     \hline
     \multicolumn{6}{l}{uniCOIL/DTHS variations.  Default: $\alpha$=0.9, $\beta$=0.2, $F_s$=$F_f$=1; Independent view} \\
     +$\alpha$=1 $\beta$=0;Single threshold  & 89(197) & 0.704 & 89.71\% & 69.00\% & 138439 \\
     +$\alpha$=1 $\beta$=0  & 84(172) & 0.704 & 89.45\%  & 68.99\% & 138353 \\
     +$\alpha$=1 $\beta$=0; Uniform & 185(390) & 0.703 & 99.99\% & 86.06\% & 521364 \\
     +$\alpha$=0.9 $\beta$=0 & 83(169) & 0.704 & 88.77\% & 68.99\% & 138283 \\
     \cline{1-1} 
     +$\alpha$=$\beta$=0.2 & 222(696) & 0.707 & 100\% & 88.93\% &  701961 \\
     +$\alpha$=0.9 $\beta$=0.2; Uniform & 170(333) & 0.707 & 99.99\% & 87.76\% & 787263 \\
     Default & 83(169) & 0.707 & 96.53\% & 69.00\% & 140719 \\
     +$F_s$=1.3 & 48(101) & 0.702 & 66.40\% & 47.31\% & 89683 \\
     +$F_s$=1.5 & 44(93) & 0.700 & 56.44\% & 42.54\% & 60485 \\
     +$F_s$=1.7 & 39(81) & 0.701 & 49.27\% & 39.53\% & 46351 \\
     \hline
    \end{tabular}
    }
\end{table}

\noindent
{\bf Result overlapping,  skipping effectiveness,  and design options.} 
Table~\ref{tab:options} compares DTHS with several baselines and impacts of its 
design options for  DL'19 passages using uniCOIL weights when $k=1,000$. 
The default DTHS setting is listed in Row 5 and each design variation listed below changes few parameters in the  default setting.
Column marked ``Overlap'' is the percentage of overlapping documents in top-$k$ compared with
the expected final ranking, which measures the relative rank-safeness. 
For BM25-T5, uniCOIL with $F=1.7$,  and DTHS variations with $\beta=0$, 
their result  overlapping ratio listed is compared against the uniCOIL final ranking results.
For other  design variations of DTHS with $\beta=0.2$, 
result overlapping  is compared against DTHS with $\alpha=\beta=0.2$, in which skipping strictly follows final rank scoring
based on a linear combination of  BM25 (weighted 0.2) and uniCOIL (weighted 0.8).

Column marked ``BLoad'' in Table~\ref{tab:options} 
is the percentage of posting blocks loaded for decompression and possible scoring. Less blocks loaded imply less  passages fully scored.
Column marked ``\# Eval'' is  the number of fully-scored passages  during retrieval. 
The above two numbers measure skipping effectiveness, and smaller such numbers have a good correlation with a faster response time
as shown by this table. This table shows that skipping  in DTHS is effective. For example, VBMW with default DTHS setting only loads 69\% of posting blocks while
unmodified VBMW for uniCOIL weights loads over 95\% of blocks.
 

In terms of DTHS design options, 
hybrid  scoring for final rank scores  visibly improves relevance by  comparing DTHS   $\alpha=\beta=0.2$ with uniCOIL ranking, and 
by comparing DTHS $\beta=0$ with $\beta=0.2$.
Hybrid scoring for  skip threshold comparison slightly improves pruning effectiveness 
by comparing ``$\alpha$=1, $\beta$=0'' and ``$\alpha$=0.9, $\beta$=0''.
Comparing Rows 6 and 7, 
dual-threshold (DT) option is about 5\% faster than the single threshold  option while their relevance is similar.
Based on the result difference between   Rows 7 and 8 and  between Rows 11 and 12,
when two queues are managed  in a uniform view, DTHS final results have  99.99\% overlapping with the expected ranking.
Thus this option  is almost rank-safe, but it reduces skipping opportunities significantly and MRT becomes much larger  than that in an independent view.



\section{Concluding Remarks} 
This paper proposes a  dual-guidance scheme (DTHS) for  document retrieval with learned sparse representations.
This scheme 
exploits both BM25 weights and learned weights compositely 
to  guide skipping with dual thresholds and improve final ranking relevance.
The evaluation shows DTHS effectively accelerates retrieval in
mean response times and 95th percentile times while delivering a very competitive  relevance.
DTHS is significantly faster than a threshold enlarging strategy in reaching a similar relevance level. 
For relatively large $k$ values, DTHS with threshold overestimation can accelerate retrieval further.
Our evaluation is reported on VBMW. The result using  BMW has  a similar pattern and  is not reported here. 
Our future work is to assess the effectiveness of dual guidance  in  other retrieval algorithms  which use
threshold-based skipping. 

\comments{
Considering that the average number of words
in queries of popular search engines is
between 2 and 3~\cite{SMH99,JS06},
the proposed technique can be very effective for a search engine which deploys
many disjoint index partitions and
when $k$ does not need to be large for each individual partition
that contributes part of top results.
}

\bibliographystyle{ACM-Reference-Format}
\normalsize
\bibliography{reference,bib/jinjin_thesis,bib/2022refer,CQbib/ranking,CQbib/mise}
\end{document}